# A Survey on Web Service Discovery Approaches

Debajyoti Mukhopadhyay, Archana Chougule

Department of Information Technology
Maharashtra Institute of Technology
Pune 411038, India
{debajyoti.mukhopadhyay, chouguleab}@gmail.com

**Abstract.** Web services are playing an important role in e-business and e-commerce applications. As web service applications are interoperable and can work on any platform, large scale distributed systems can be developed easily using web services. Finding most suitable web service from vast collection of web services is very crucial for successful execution of applications. Traditional web service discovery approach is a keyword based search using UDDI. Various other approaches for discovering web services are also available. Some of the discovery approaches are syntax based while other are semantic based. Having system for service discovery which can work automatically is also the concern of service discovery approaches. As these approaches are different, one solution may be better than another depending on requirements. Selecting a specific service discovery system is a hard task. In this paper, we give an overview of different approaches for web service discovery described in literature. We present a survey of how these approaches differ from each other.

## 1 Introduction

Web services are application components which are based on XML [2]. Web services can be used by any application irrespective of platform in which it is developed. Web service description is provided in WSDL document. It can be accessed from internet using SOAP protocol. In industry, many applications are built by calling different web services available on internet. These applications are highly dependent on discovering correct and efficient web service. The discovered web service must match with the input, output, preconditions and effects specified by the user. Even after functional matching, QoS parameters also need to be matched to have best web service from available web services. Web services developed by different vendors are published on internet using UDDI [1]. UDDI is the mechanism for registering and discovering web services. It is platform independent registry as it is based on extensible markup language. It allows businesses to give list of services and describe how they interact with each other. In literature, many approaches for web service discovery are described some of which work on UDDI. Search in UDDI is based on keyword matching which is not efficient as huge number of web services may match a keyword and it is difficult to find the best one. Other approaches take advantage of semantic web concept where web service matching is done using ontologies. Discovering web services automatically without human interface is an important concern. Different approaches to for automatic discovery of web services are also suggested by authors. This paper is organized as follows: section 2 gives overview of web service discovery process. Section 3 describes service discovery approaches. We conclude the paper in section 4.

## 2   Web Service Discovery

A web service discovery process is carried out in three major steps. First step is advertisement of web service by developers. Providers advertise web services in public repositories by registering their web services using web service description file written in WSDL [3]. Second step is web service request by user. User sends web service request specifying the requirement in predefined format to web service repository. Web service matcher which is core part of web service discovery model, matches user request with available web services and finds a set of web service candidates. Final step is selection and invocation of one of the retrieved web services. Discovery of correct web service depends on how mature web service matching process is. i.e.; how actual requirements of user are represented in formalized way and how they are matched with available services.

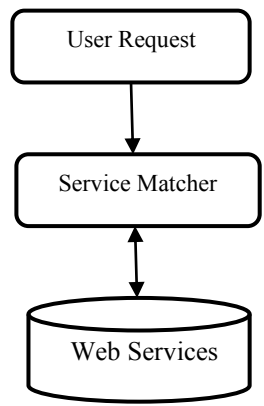

**Fig. 1.** Web service discovery

## 3   Survey

In this section we give overview of thirteen different approaches for web service discovery. For each one, we mention the details where one approach differs from others.

### 3.1   Context aware web service discovery

As format for sending web service request is fixed, some information in user's request is lost during transforming user's request to formalized one. To overcome this limitation, context aware web service discovery approach is suggested by Wenge Rong and Kecheng Liu [4]. Context aware discovery is useful for request optimization, result optimization and personalization. As concept of context is very complex, they suggest with an example that context should be domain oriented or problem oriented. The context in web service discovery is formally defined as any information that explicitly and implicitly affects the user's web service request generation. They divide context in two categories as Explicit and implicit. Explicit context is directly provided by the user during matchmaking process such as Q&A information. Implicit context is collected in automatic or semi-automatic manner. Implicit context is more applicable to web service discovery as user is not directly involved. Context awareness is again divided in four categories depending on how context is collected. The categories are Personal profile oriented context, Usage history oriented context, Process oriented context and other context. Personal profile oriented context is collected using user's personal profile which contains personal data, preferences and other information. Personalization information such as location, time and user's situation is used for decomposing the discovery goal, setting selection criteria and supplying parameters. Limitation of this method is, it makes system architecture more complicated when new attributes and constraints are introduced. Usage history oriented context is collected for predicting user's next behavior. It is based on assumption that web service requests by specific user are similar during a certain period of time. Usage history oriented context is again divided in two categories as Personal usage history oriented context and Group usage history oriented context. User's previous system interaction can be stored in system log. Log records can be used to provide recommendation for service selection decision. But the user may not have similar requirements afterwards. So Group usage history oriented context is used where web service matchmaking is based on behavior information of other user groups in similar situation. One of the examples of Group oriented context awareness is collaborative filtering (CF) which may be memory based or content based. Group oriented context can also be collected from observation data in particular community. Process oriented context is built from user sessions. User reactions to retrieved web services are understood in particular request session and the discovery is optimized. This feed based process oriented context can be built using Probabilistic Latent Semantic Analysis (PLSA) [5]. In case where single web service is not sufficient to complete user request, composition of multiple web services is carried out. In this case, context should be built considering composite web service discovery process. This approach is better than traditional keyword matching used in UDDI as user intension is understood better.

**3.2 Publish subscribe model**

Falak Nawz, Kamram Qadir and H. Farooq Ahmad[6] propose push model for web service discovery where service requesters are provided with service notification prior to discovery. They use semantic based web service matching where service descriptions are matched using OWL-S [7], an ontology language for web service description. They also rank published web services depending on the scores assigned using concept matching. They divide the system in two phases as subscription phase, which starts when a subscriber registers himself onto registry for notification of required services and notification phase, which starts when a new service is published on registry. In subscription phase, when user goes for subscribing, subscription information along with his/her location and specific web service requirements are stored in subscription knowledge base. Information from knowledge base is used later for service matching. Information in knowledge base is stored in OWL format. Service categories are maintained according to user requests received till date. In each service category, lists are maintained containing information about number of input and output parameters required by each subscription. The best matching web service is selected by matching user requirements (inputs, outputs, preconditions and effects) to OWL-S descriptions stored in registry. Matching can be in one of six levels as Exact, Plug-In, Subsume, Enclosure, Unknown and Fail. In notification phase, OWL-S descriptions for newly registered web services are added to matching subscription categories and listing for number of parameters is updated. If there is no single matching subscription category, then service descriptions are directly added to registry. Subscriptions in knowledge base are stored in the form of subsumption relationship ontology. Subscribers are notified when their leasing subscription time expires so that subscribers can renew their subscriptions. Time required for web service discovery is minimized with this approach as search area is reduced to specific category. Probability of finding most suitable web service also increases. Limitation of this approach is, it adds overhead in developing and maintaining new components in system architecture.

### 3.3 Keyword clustering

Web service discovery based on Keyword clustering and concept expansion is suggested by J. Zhou, T. Zhang, H. Meng, L. Xiao, G. Chen and D. Li[8]. They calculate similarity matrix of words in domain ontology based on Pareto principal and use that for semantic reasoning to find matching service. Bipartite graphs are used to find matching degree between service requests and available services. They describe in detail how Kuhn-munkres algorithm can be used to compute optimal matching of a bipartite graph.

### 3.4 Service request expansion

One more approach for enhancing web service discovery is sending modifying user requests as suggested by A. Paliwal, N. Adam and C. Bornhovd [9]. They expand service requests by combining ontologies and latent semantic indexing. They build the service request vector according to the domain ontology, build the training set of the LSI classifier by extracting features from selected WSDL files, and then project the description vectors and the request vector. They utilize the cosine measure to determine similarities and to retrieve relevant WSDL service descriptions. Ontology linking is done using semi automated approach. It is done by mapping domain ontologies to upper merged and mid-level ontologies. Keywords are selected from service request by pre-processing service request which includes removal of mark-ups, punctuations and use of white spaces etc. Keyword based search is applied to upper ontology and relevant ontology is identified from ontology framework. Service request is expanded by acquiring associated concepts related to initial service request with semantic matching and assembling of concepts and enhanced service request is achieved. From collection of WSDL documents, relevant WSDL documents are found and service description set is built. Service description set is then transformed into a term-document matrix by parsing and processing of documents. Removal of mark-ups and index entries, removal of punctuations, stoplist, use of white space as term delimiters and stemming to strip word endings are the steps involved in WSDL document processing. Term-document matrix is generated out of WSDL processing which indicates term frequencies. Built training set is used for LSI. LSI includes Singular Value Decomposition (SVD). In SVD, original matrix is approximated by a linear combination of a decomposition set of term to text-object association data. The resulting description vectors and request vectors are then projected and similarity is calculated using cosine similarity. At last, resulting web services are ranked based on similarity measure. Disadvantage of this approach is cost of computing LSI and string SVD is high.

### 3.5 BPEL processes ranking using graph matching

When user requests for web service in available web services repository if exact matching web service does not exist, then approximate matching web service can be suggested by service matcher. To achieve this goal, behavioral matching is required. D. Grigori, J. Carlos Corrales, M. Bouzeghoub and A. Gate [10] developed matching technique which works on BPEL [11] behaviour model. User requirements are expressed as a service behaviour model. They transform BPEL specification to a behaviour graph using flattening strategy and transform service matching problem to graph matching problem. In graphs, regular nodes represent the activities and connectors represent split and join rules. Flattening strategy maps structural activities to respective BPEL graph fragments. The algorithm traverses the nested structure of BPEL control flow (BCF) in a top-down manner and applies recursively a transformation procedure specific to each type of structured activity. This procedure checks whether the current activity serves as target and source for links and adds arcs or respective join and split connectors in the resulting graph fragment. Five structural activities handled are Sequence, Flow, Switch, While and Pick. The generated process graph which represents user requirements is then compared with the target graphs in library. Error correcting graph matching is used to find approximate matching process model if exact matching process graph is not available. Similarity is measured as inverse of distance between two graphs representing BPEL. Distance is defined as cost of transformations needed to adapt the target graph in order to cover a subgraph in the request graph. Different measures for calculating cost are defined as distance between two basic BPEL activities, matching links between connector nodes and linguistic similarity between two labels based on their names. The results are optimized by applying granularity-level analyzer. It checks whether composition/decomposition operations are necessary for graph matching. BPEL processes are then ranked in decreasing order of calculated distance between graphs and web services. The limitation of this approach is method is completely based on syntactic matching. Semantics of user request is not considered.

**3.6 Layer based semantic web service discovery**

Finding a matching web service in whole service repository is time consuming process. Guo Wen-yue, Qu Hai-cheng and Chen Hong [12] have divided search in three layers by applying filters at each layer and thus minimizing search area. They have applied this approach on intelligent automotive manufacturing system. Three layers for service matching are service category matching, service functionality matching and quality of service matching. Semantic web service discovery is done based on OWL-S, using ServiceProfile documents for service matching. First step in service discovery is service category matching. Service category matching is carried out to minimize time and storage space required for service matching. At this layer, service category matching degree is computed. ServiceCatogory attribute in ServiceProfile contains category of service. This value is matched against service category of request which is passed b user while sending request. If there is match, web service is selected to enter the next service functionality matching layer. Advertisements that do not meet the demands are filtered out. Then service functionality matching degree is computed in the service functionality matching layer. For functionality matching, four attributes defined in ServiceProfile are matched against service request. These attributes are hasInput, hasOutput, hasPrecondition and hasResult. Advertisements that do not meet the conditions are filtered out, while other advertisements that satisfy the conditions are selected to enter the next quality of service matching layer. Last step is computing quality of service matching degree. Quality of service is decided based on response time of service discovery and reliability of service discovery system. From service category matching degree, service functionality matching degree and quality of service matching degree, service matching degree is calculated and the advertisements that best meet needs of requesters are presented to requesters in the form of list.

**3.7 Service discovery in heterogeneous networks**

Web services are heavily used by military networks which are heterogeneous and decentralized in nature. There is need of interoperable service discovery mechanism to enable web service based applications in military networks. Traditional mechanisms for web service discovery such as UDDI and ebXML are not suitable in military networks as they are centralized and cannot be available during network partitioning.. F. Johnsen, T. Hafsoe, A. Eggen, C. Griwodz and P. Halvorsen[13] suggest the web service discovery solution which can fulfil the requirements in military networks. As same protocol cannot be used in heterogeneous networks, they suggest using of service discovery gateways, so that each network domain can employ the most suitable protocol. Interoperability is ensured by using service discovery gateways between the domains that can translate between the different service discovery mechanisms. Creation and interpretation of service descriptions in clients, servers, and gateways are done to ensure interoperability. This mechanism is called as Service Advertisements in MANETs (SAM), a fully decentralized application-level solution for web services discovery. It integrates periodic service advertisements, caching, location information, piggybacking, and compression in order to be resource efficient. A gateway periodically queries all services in the WS-Discovery and proprietary domains. Services that are available (if any) must then be looked up in the gateway's local service cache. This local cache is used to distinguish between services that have been discovered, converted, and published before, and new services that have recently appeared in each domain. If a service is already present in the cache, it has been converted and published before, and nothing needs to be done. On the other hand, if the service is not in the cache, it is translated from one service description to the other, published in the network, and added to the local cache. For each query iteration, gateway compares local cache containing all previously found services with the list of services found now. Service is removed from the local cache if it deleted from its domain. This behaviour allows the gateway to mirror active services from one domain to the other, and remove any outdated information. They have implemented a gateway prototype solving transparent interoperability between WS-Discovery and a cross-layer solution, and also between WS-Discovery and SAM.

### 3.8 Web service indexing

To enable fast discovery of web services, available web services can be indexed using one of the indexing mechanisms such as inverted indexing and latent semantic indexing. B. Zhou, T. Huan, J. Liu and Meizhou Shen [14] describe how inverted indexing can be used for quick, accurate and efficient web service discovery. In semantic web service discovery, user request is matched against OWL-S descriptions of web services. In this case, inverted index can be used to check whether the OWL-S description with the given id contains the term. Inverted index consists of list of keywords and frequency of keyword in all OWL-S documents. Every keyword is connected to a list of document ids in which that keyword occurs. They have suggested extensions to inverted lists to find positions of terms in OWL-S descriptions.

M. Aiello, C. Platzer, F. Rosenberg, H. Tran, M. Vasko and S. Dustdar[15] describe VitaLab system which is web service discovery system based on indexing using hashtable. They have implemented indexing on WSDL descriptions which are parsed using Streaming API for XML (StAX). Two hash tables namely parameter index and service index are built. Parameter table maintains the mapping from each message into two lists of service names for request and response respectively, to get a list of services that consume or produce a particular message. Service index maps service names to their corresponding detail descriptions. Generated indexes are serialized as binary files and stored in non-volatile memory and used the same every time when new service is added or existing service is modified or deleted.

One more index structure for concept based web service discovery is used by C. Wu, E. Chang and A. Aitken[16]. They use Vector Space Model (VSM) [17] indexes and Latent Semantic Analysis (LSA) indexer on term document matrices generated by processing WSDL descriptions which are retrieved by web crawlers. Term document matrices are generated according to Zipf Law and by applying Singular Value Decomposition (SVD). VSM indexer takes all term documents as input and outputs WSDL indices representing term document matrices. These term matrices are given as input to LSA indexer which generates as output semantic space for service retrieval.

Advantage of indexing approach is, once the indexes are available, it is easy to retrieve the objects fast using index. Limitation of the approach is, indexing process is computationally expensive. It requires additional space to store the indexes and the indexes need constant update if data changes often.

### 3.9 Structural case based reasoning

Georgios Meditskos and Nick Bassiliades[18] describe semantic web service discovery framework using OWL-S. They detail a web service matchmaking algorithm which extends object-based matching techniques used in Structural Case-based Reasoning. It allows retrieval of web services not only based on subsumption relationships, but also using the structural information of OWL ontologies. Structural case based reasoning done on web service profiles provide classification of web services, which allows domain dependent discovery. Service matchmaking is performed on Profile instances which are represented as objects considering domain ontologies. In Semantic case based reasoning (SCBR), similarity is measured as interclass similarity considering hierarchical relationships and intraclass similarities by comparing attribute values of objects of same class. Web service discovery is done by measuring similarity at three levels as taxonomical similarity, functional similarity and non-functional similarity. Taxonomical similarity between advertisements and query is the similarity of their taxonomical categorization in a Profile subclass hierarchy. It is calculated using DLH metric which represents the similarity of two ontology concepts in terms of their hierarchical position. Four hierarchical filters for matching are defined as exact, plugin, subsume and sibling. Functional similarity is calculated based on input and output similarity (signature matching) of advertisement and query. It is ensured whether all the advertisement inputs are satisfied by query input and all query outputs are satisfied by advertisement outputs. Non-functional similarity is measured by directly comparing values of data types and objects. They calculate overall similarity between advertisement and query in terms of their taxonomical, functional and non-functional similarity. The semantic web service discovery framework is further enhanced to perform service discovery using ontology roles as annotation constraints. The framework is implemented using OWLS-SLR [19] and compared with OWLS-MX matchmaker.

### 3.10 Agent based discovery considering QoS

As there can be multiple web services available providing same kind of functionality, best service among them should be selected. This can be done using QoS parameters. T. Rajendran and P. Balasubramanie[20] suggest a web service discovery framework consisting of separate agent for ranking web services based on QoS certificates achieved from service publishers. Main entity of web service discovery framework is verifier and certifier which verifies and certifies QoS of published web service. The service publisher component is responsible for registration, updating and deletion of web service related information in UDDI. Service publisher is supplied with business specific and performance specific QoS property values of web services by service providers. Verification and certification of these properties is then done by web service discovery agent. After that, service provider publishes its service functionality to UDDI registry through service publisher. The service consumer searches UDDI registry for a specific service through discovery agent which helps to find best quality service from available services which satisfies QoS constraints and preferences of requesters. QoS verification is the process of validating the correctness of information described in service interface. Before binding the web service, service consumer verifies the advertised QoS through the discovery agent. The result of verification is used as input for certification process. Backup of certificates is also stored by web service agent which is used for future requests for similar kind of web services. Time required for selecting web service with best QoS values eventually decreases. The QoS parameters for selecting best web service are suggested by authors. These parameters are response time, availability, throughput and time. Values of these parameters are stored in tModels of respective web services which are supplied by service publisher.

**3.11 Collaborative tagging system**

In feedback based web service discovery, comments from users who already have used the web services can be useful for other users. This approach is adapted by U. Chukmol, A. Benharkat and Y. Amghar[21]. They propose collaborative tagging system for web service discovery. Tags are labels that a user can associate to a specific web service. Web services are tagged by different keywords provided by different users. For each tag, tag weight is assigned. A tag weight is the count of number of occurrences of a specific tag associated to a web service. Tag collection is the collection of all entered tags. Each tag in tag collection is associated to a certain number of web services, forming resource vector. They employ both types of tags as keyword tag and free text tags. These tags are made visible to all users who want to access web services. When user sends a query for keyword based discovery, matching web service is found by checking whether there exists a web service having tag matching exactly with user input. If not, it checks whether synonym to the keyword exists with help of synonym set obtained from WordNet. System provides support for preparing queries using AND, OR and NOT operation. User can also attach more than one keyword as tag. This is called as free text tagging. In this case, each web service is associated with multiple keyword tags. Keywords are arranged one below another called as aggregated text tag. It is then transformed in vector of terms. Service discovery using free text is also provided. When user sends query as free text, it is converted into vector of terms to find matching web service. Vector Space Model is employed to carry out vector term matching. They use the Porter stemming algorithm [22] to extract terms vector from a query document and the aggregated text tag associated to web service. Similarity is calculated as is the cosine value between the two vectors representing both texts. Resulted web services are ranked according to the values of cosine coefficient between the query text and all aggregated text tags associated to resources.

**3.12 Peer-to-Peer discovery using finite automaton**

As centralised web service discovery approach has many disadvantages such as single point of failure and high maintenance cost, F. Emekci, O. Sahin, D. Agrawal and A. Abbadi[23] propose a peer-to-peer framework for web service discovery which is based on process behaviour. Framework considers how service functionality is served. All available web services are represented using finite automaton. Each web service is defined as follows: A Web service p is a triple, p=(I, S, R), such that, I is the implementation of p represented as a finite automaton, S is the service finite automaton, and R is the set of request finite automata. When user wants to search for web service, PFA of finite automaton of web service(R) is sent for matching. Matching is done against S by hashing the finite automata onto a Chord ring. Chord is a peer-to- peer system for routing a query on hops using distributed hash table. Regular expression of the queried PFA is used as the key to route the query to the peer responsible for that PFA.

**3.13 Hybrid approach**

Main categories of web service discovery approaches are keyword based and ontology based. Y. TSAI, San-Yih, HWANG and Y. TANG [24] make use of both approaches for finding matching web service. The approach considers service providers information, service descriptions by providers, service description by users, operation description by providers, tags and categories and also QoS attributes. For finding similarity between query and candidate web service, similarity between two operations is calculated first. For this, similarity between input/output of query and web service is calculated using ontology of web service. For a given operation input (output), each of its message part is mapped to a concept in an ontology. This similarity is tested at three levels. First relative positions of the concepts associated to query and web service message parts are considered. The position can be one of three parts, 1) exact where two classes are same, 2) subsume where one concept is super class of other class and 3) others where two concepts are not related. At second layer, similarity is measured based on paths between two concepts and at last similarity is measured based on information content (IC) of concepts. After considering similarity between input/output attributes, other attributes such as name, description, tags and operation name are considered. For these attributes, text based method is used where two words are compared lexically measuring the longest continuous characters in common. For measuring overall similarity, weights are assigned to all the attributes using analytical hierarchy process. AHP is the method for multiple criteria decision making [25]. Overall similarity between query and web service is calculated as weighted sum of similarities of all associated attributes. Described approach is tested and compared with text based approach and ontology based approach and it is shown that, hybrid approach gives better results than using each approach separately.

## 4   Conclusion

Success of published web services depends on how it is getting discovered. Efficiency, accuracy and security factors must be considered while providing discovery mechanism. We have given overview of different web service discovery approaches with their advantages and disadvantages. Many approaches differ in the way web service matching is carried out. Some approaches are considering concept of semantic web, while some other focus on information retrieval methods. Some approaches suggest enhancement in web service request based on metadata about web services generated by feedback of other users. Some approaches suggest additional tools in traditional framework of web service discovery. Minimizing total search area using clustering techniques is also suggested. Survey shows that considering QoS parameters while selecting is important because, number of available web services providing same kind of functionality is very large. As web service discovery requiring manual interference may take more time, solutions for automatic discovery are drawing more attention.